\documentclass[%
 preprint,
superscriptaddress,
 amsmath,amssymb,
 aps, physrev,
]{revtex4-2}

\usepackage{graphicx}
\usepackage{dcolumn}
\usepackage{bm}
\usepackage{hyperref}

\usepackage[usenames]{color}
\usepackage{setspace}
\renewcommand{\footnoterule}{%
  \hrule width \textwidth height 1pt
  \kern 2pt
}

\begin{document}

\title{\textbf{A Single-granule Stirling Heat Engine} }

\author{Niloyendu Roy}
\altaffiliation[Presently at ]{Department of Physics, University of Konstanz, 78457 Konstanz, Germany}
\affiliation{Chemistry and Physics of Materials Unit, Jawaharlal Nehru Centre for Advanced Scientific Research, Jakkur, Bengaluru - 560064, INDIA}
\author{Pragya Arora}
\affiliation{Chemistry and Physics of Materials Unit, Jawaharlal Nehru Centre for Advanced Scientific Research, Jakkur, Bengaluru - 560064, INDIA}

\author{A K Sood}
\affiliation{Department of Physics, Indian Institute of Science, Bangalore 560012, INDIA
}
\affiliation{International Centre for Materials Science, Jawaharlal Nehru Centre for Advanced Scientific Research, Jakkur, Bengaluru - 560064, INDIA
}
\author{Rajesh Ganapathy}
\affiliation{Chemistry and Physics of Materials Unit, Jawaharlal Nehru Centre for Advanced Scientific Research, Jakkur, Bengaluru - 560064, INDIA
}
\affiliation{International Centre for Materials Science, Jawaharlal Nehru Centre for Advanced Scientific Research, Jakkur, Bengaluru - 560064, INDIA
}
\affiliation{School of Advanced Materials (SAMat), Jawaharlal Nehru Centre for Advanced Scientific Research, Jakkur, Bengaluru - 560064, INDIA
}
\email{Contact author: niloycrj@gmail.com}
\email{Contact author: rajeshg@jncasr.ac.in}
\date{\today}
\begin{abstract}

\textbf{Single-particle heat engines at atomic and colloidal scales obey the universal thermodynamic bounds on work and efficiency. Here, we translate these principles to the macroscale by building an \textit{athermal} Stirling engine whose working medium is a millimeter-sized, vibrofluidized granule confined in a time-dependent magnetic trap. By embedding a `rattler' within the granule to inject noise, we engineer overdamped, Brownian-like dynamics in an otherwise inertial particle. This design enables independent control over the granule’s effective temperature and spatial confinement. Our engine quantitatively reproduces the universal power-efficiency trade-offs of finite-time thermodynamics, achieving the Curzon-Ahlborn efficiency at maximum power. Strikingly, we uncover a control parameter-dependent damping that leads to an unexpected dissipation mechanism - the losses in the compression stroke rival or even exceed those during expansion. Our work establishes an accessible experimental platform to study small-system thermodynamics in intrinsically athermal systems.} 
\end{abstract}

\maketitle
\newpage
Single-particle heat engines provide fundamental insights into the thermodynamics of small systems \cite{schmiedl2007efficiency,seifert2012stochastic,van2013stochastic,martinez2017colloidal}, much as their macroscopic counterparts helped establish the foundations of classical thermodynamics \cite{carnot1890reflections,callen2006thermodynamics}. The engines built so far utilize only a single atom \cite{rossnagel2016single} or colloid as their working medium \cite{blickle2012realization}, making their energetics highly sensitive to environmental fluctuations. This unique sensitivity allows these engines to operate not just across conventional thermal reservoirs \cite{blickle2012realization,martinez2016brownian}, but also across non-equilibrium baths (e.g., active \cite{krishnamurthy2016micrometre}, non-Markovian \cite{krishnamurthy2023overcoming}), thereby revealing the core principles of heat-to-work conversion in complex environments \cite{zakine2017stochastic,kumari2020stochastic,holubec2020active,roy2021tuning,fodor2021active,gronchi2021optimization,krishnamurthy2022synergistic,roy2023harnessing,chang2023stochastic}. 

Here, we extend this principle into a new regime by constructing a Stirling heat engine with a millimeter-sized, vibrofluidized granule as its working medium. Unlike atomic or colloidal systems, driven granules are intrinsically out of equilibrium because their collisions are inelastic, constantly dissipating energy \cite{goldhirsch2008introduction,puglisi2014transport}. Yet, an effective equilibrium description can emerge when these systems are held in a non-equilibrium steady state, \cite{d2003observing,feitosa2004fluidized,ojha2004statistical,song2005experimental}, enabling the operation of Brownian-like ratchets \cite{eshuis2010experimental,joubaud2012fluctuation,gnoli2013brownian}, and Maxwell's Demon \cite{lagoin2022human}. However, these apparent parallels to equilibrium are an exception, not the rule \cite{kudrolli1997cluster, olafsen1999velocity,falcon1999cluster,opsomer2012dynamical}. Building a single-granule heat engine, therefore, presents a formidable task because it demands precise and cyclical control over system parameters, such as its effective temperature and volume -- a feat rarely achieved in granular systems. 

We chose the Stirling cycle \cite{blickle2012realization} as it avoids the complexities involved with achieving adiabatic isolation of the particle from its environment required by the more familiar Carnot cycle \cite{martinez2016brownian}. Furthermore, operating the engine in the overdamped limit simplified its energetics by mitigating inertial effects \cite{schmiedl2007efficiency,tu2014stochastic,zoller2017optimization}. In a standard Stirling engine, the working medium undergoes expansion in the hot reservoir and compression in the cold one, with these isothermal steps linked by isochoric heating and cooling. To replicate this cycle with a single particle, we must simultaneously tune the effective bath temperature, $T$, and the stiffness of the confining potential, $k$ --- the mesoscale (small system size) analog of the piston. Independent control over $T$ and $k$ would trace a rectangle in the $T-k$ parameter space in equilibrium conditions \cite{krishnamurthy2023overcoming}. Our immediate goal, therefore, was to achieve this level of control with a single granule.  

\begin{figure}[htbp]
\includegraphics[width=1\textwidth]{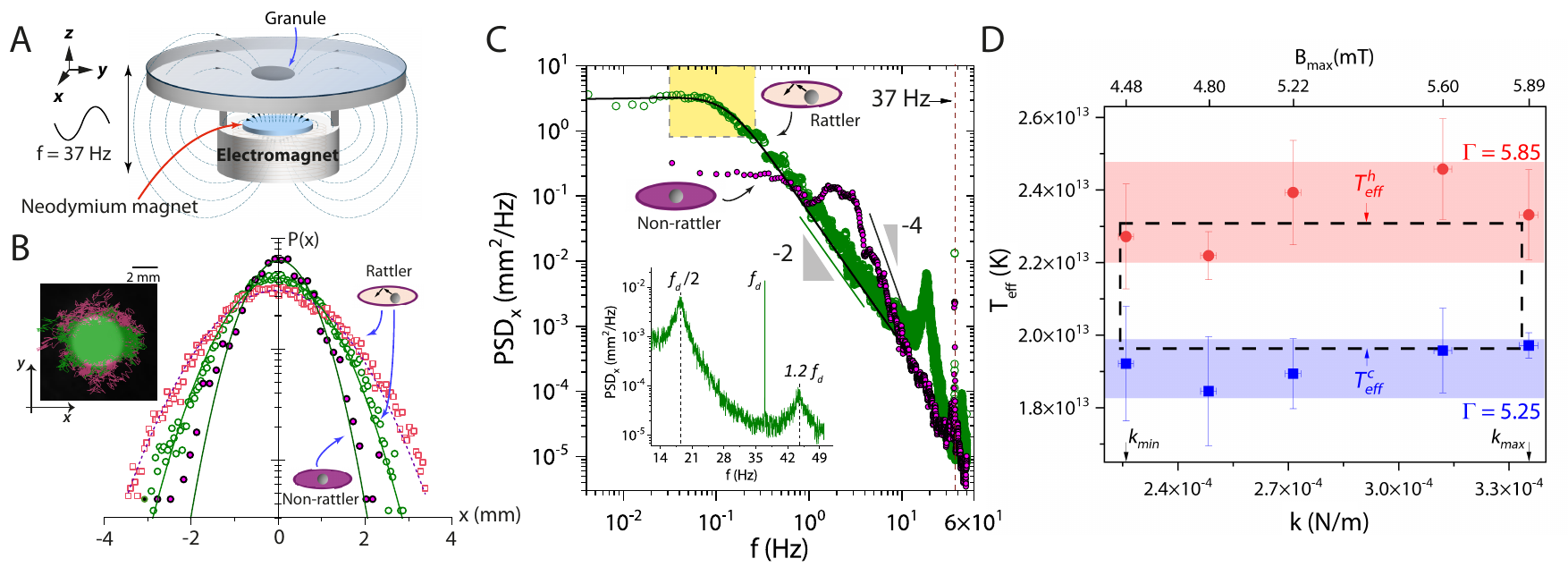}
\caption{\textbf{Engineered Brownian-like dynamics of a granule and realization of a $T_{eff}-k$ diagram.} \textbf{(A)} Schematic of the experimental set-up. The working medium is an oblate spheroidal granule made by gluing two 3D-printed plastic shells enclosing a weakly ferromagnetic ball bearing. The Neodymium magnet creates the primary trap whose strength is varied by modulating the current through an additional electromagnet. The injected energy is tuned via the vibration amplitude. We designed two types of particles: a `non-rattler', where the ball-bearing (1.5 mm in diameter) is fixed within the granule, and a `rattler', where the ball bearing (1 mm or 1.5 mm in diameter) is free to rattle. \textbf{(B)} Position probability distribution along the $x-$axis, $P(x)$, for the non-rattler (filled circles) and smaller (1 mm) rattler (hollow symbols) for dimensionless acceleration, $\Gamma = 5.85$. For the rattler, data are presented at both the minimum ($k = k_{min} = 2.25\times10^{-4}$ N/m, hollow squares) and maximum ($k = k_{max} = 3.35\times10^{-4}$ N/m, hollow circles) trap stiffness. The particle trajectory is overlaid on the image for the two $k$ values (left panel) \textbf{(C)} Power spectral density, $PSD_x$, for the data shown in \textbf{(B)}. There is a sharp peak at the energy injection frequency $f_d$. The black line is a Lorentzian fit to the data. The inset is an expanded view of the spectra for the rattler. The yellow shaded region represents the operating frequencies of the Stirling cycle (Fig. \ref{Figure2}A). \textbf{(D)} $T_{eff}-k$ diagram for the smaller rattler. The dashed line represents the best-fit rectangle to the data. All error bars denote the standard error of the mean (SEM).}
\label{Figure1}
\end{figure}

The working medium is an oblate granule (4 mm major-axis and 2.1 mm minor-axis) enclosing a weakly ferromagnetic steel ball bearing (1 mm or 1.5 mm in diameter) (Fig. \ref{Figure1}A). We held the granule in a magnetic trap, and the entire assembly was subject to vertical vibrations from a shaker at a frequency $f_d$ kept fixed at 37 Hz (see Methods). The dimensionless acceleration, $\Gamma = \frac{4\pi^2 af_d^2}{g}$, a measure of the injected energy \cite{d2003observing}, was varied between $5.25\leq\Gamma\leq5.85$ by changing the vibration amplitude $a$. Here, $g$ is the acceleration due to gravity. We confined the granule from above with a glass plate, which restricted its dynamics to be quasi-two-dimensional and permitted imaging. A permanent magnet and an electromagnet together create a tunable harmonic confining potential, $U(x) \approx \frac{1}{2}kx^2$, allowing us to measure the trap stiffness precisely (see SI Text and Supplementary Figs. 1-5). The data shown in the main manuscript, unless explicitly stated, are for the smaller 1 mm rattler. Data for the 1.5 mm rattler are provided in the SI. 

Our initial experiments with the ``non-rattler" particle, where the internal ball bearing was immobilized, highlighted the intrinsically inertial nature of the granule. While the particle position distribution, $P(x)$, deviated marginally from a Gaussian expected for a harmonically bound particle (Fig. \ref{Figure1}B), the power spectral density, $PSD_x(f)$, showed the characteristic $f^{-4}$ scaling of underdamped dynamics beyond the trap roll-off frequency $f_c$ (Fig. \ref{Figure1}C) \cite{kheifets2014observation,bian2016111}.  To suppress the inertial dynamics, we allowed the ball bearing to rattle within the granule, causing it to act as an additional source of noise. The results were striking. The $PSD_x$ for the ``rattler'' became Lorentzian and scaled as $f^{-2}$ for over two decades beyond $f_c$ (Fig. \ref{Figure1}C), consistent with overdamped Brownian dynamics \cite{bian2016111}. Concurrently, $P(x)$ is well-fit by a Gaussian, and it also widened predictably as $k$ decreased, confirming that we could modulate the effective volume available to the granule (Fig. \ref{Figure1}B, also see Supplementary Fig. 6).

This shift to overdamped behaviour stems from the internal dynamics of the rattler. The $PSD$ reveals, in addition to the sharp peak at $f_d$, new broad peaks at $f_d/2$ and 1.2$f_d$ (inset to Fig. \ref{Figure1}C). These spectral features correspond to period-doubling and quasiperiodicity, respectively, and signal that the in-plane motion of the granule shares the hallmarks of a non-linear dynamical system en route to chaos \cite{peitgen2004chaos}. This complex dynamics, when coarse-grained over longer timescales, mimics thermal noise and helps suppress inertial correlations. Importantly, these features were insensitive to the tested ranges of $\Gamma$ and $k$ (Supplementary Fig. 7). Our observations are distinct from previous studies on a shaken granule that observed chaotic dynamics along the vibration axis \cite{holmes1982dynamics,pieranski1983jumping}.

The Brownian-like behaviour justified defining an effective temperature using the equipartition theorem $\frac{1}{2}k\langle x^2\rangle = \frac{1}{2}k_BT_{eff}$. Doing so enabled testing whether we had independent control over $T_{eff}$ and $k$. As expected, increasing $\Gamma$ raised the effective temperature \cite{d2003observing} Fig. \ref{Figure1}D). Crucially, $T_{eff}$ remained essentially unchanged even as we varied $k$. This decoupling, a feature of an equilibrium-like system, allowed us to realize a rectangular $T-k$ diagram with effective cold and hot isotherm temperatures of $T^{eff}_c = 1.93\times 10^{13}$ K and $T^{eff}_h = 2.3\times 10^{13}$ K, respectively. These large effective temperatures simply reflect the high kinetic energy of the driven granule \cite{d2003observing,goldhirsch2008introduction,puglisi2014transport}. Since we constructed this diagram from steady-state measurements (fixed values of $\Gamma$ and $k$), its area represents the maximum work obtainable from the cycle, which can be achieved by an engine with cycle duration $\tau\to\infty$ (quasistatic limit) (data for the larger rattler shown in Supplementary Figs. 8-10). 

We executed a Stirling cycle by the simultaneous time-modulation of $T_{eff}$ and $k$ as per the protocol detailed in Fig. \ref{Figure2}A (also see Supplementary Fig. 5). We varied the cycle duration from 2 s to 32 s, corresponding to operating frequencies that straddle $f_c$ (shaded region in Fig. \ref{Figure1}C, see SI Text for details on the larger rattler). Importantly, the highest operating frequency (0.5 Hz) is well below $f_D$, ensuring that the particle dynamics are fully overdamped.

Stochastic thermodynamics provides a framework for quantifying the engine energetics from particle position trajectories \cite{sekimoto1998langevin,schmiedl2007efficiency,seifert2012stochastic}. The work done per cycle, $W_{cyc}$, is the change in potential energy due to trap stiffness modulation and for the $i^{\text{th}}$ cycle, $W_{cyc}=\int_{t_i}^{t_i+\tau}\frac{dU}{dk}\circ dk$, while the heat dissipated, $Q_{cyc} = \int_{t_i}^{t_i+\tau}\frac{dU}{dx}\dot{x}dt$. The $\circ$ denotes a Stratonovich product and $t_i$ marks the start of the $i^{\text{th}}$ cycle. For harmonic confinement these quantities reduce to, $W_{cyc} =\int_{t_i}^{t_i+\tau}x^2\circ dk$ and $Q_{cyc}=\int_{t_i}^{t_i+\tau}kx\dot{x}dt$. $dk$ is zero during the isochoric transitions, and the work is extracted only during the isothermal processes. Typical of a stochastic heat engine, $W_{cyc}$ shows large cycle-to-cycle variations (see Fig. \ref{Figure2}B for $\tau = 32$s, also see Supplementary Fig. 11). The corresponding work distribution, $P(W_{cyc})$, is a Gaussian with the mean work, $\langle W\rangle$, being negative, which, as per convention, implies the engine does work on its surroundings.

The cycle time dependence of $\langle W\rangle$ offered crucial insights into the engine's operation (Fig. \ref{Figure2}C). In the large $\tau$ limit, the engine performs useful ($-$ve) work; $\langle W\rangle$ saturates and the experimentally calculated work agrees well with the predicted equilibrium Stirling work for an overdamped Brownian particle, $W_{\infty} = k_B(T^{eff}_c - T^{eff}_h)\ln\sqrt{k_{max}/k_{min}}$ \cite{blickle2012realization}. Faster cycles introduce irreversible losses because the system cannot equilibrate with the rapid changes in the control parameter. This dissipation reduces the work output, eventually stalling the engine ($\langle W\rangle>0$) for the larger rattler. The impact of irreversibility on $\langle W(\tau)\rangle$, can be modeled as: 
\begin{equation}
    \langle W(\tau)\rangle = \langle W_{diss}\rangle + \langle W_\infty \rangle
    \label{W_finite}
\end{equation}
Here, to leading order, the dissipative work is inversely proportional to the cycle time, $\langle W_{diss}\rangle = \frac{\Sigma}{\tau}$ \cite{schmiedl2007efficiency,esposito2010efficiency}, where $\Sigma$ is a protocol-dependent irreversibility parameter. This model fits our data well, yielding $\Sigma$ values of $0.07k_BT^{eff}_c$ and $0.15k_BT^{eff}_c$ for the smaller and larger rattlers, respectively.

The power $\mathcal{P} = -\frac{\langle W\rangle}{\tau}$, and the efficiency, $\epsilon = -\frac{\langle W\rangle}{Q_H}$, obey the universal power-efficiency trade-off that constrains all heat engines (Fig. \ref{Figure2}D and E); overcoming this bound often requires bespoke reservoir engineering strategies \cite{klaers2017squeezed,krishnamurthy2023overcoming}. Here, $Q_H$ is the sum of heat absorbed during isochoric heating and isothermal expansion steps. When $\tau\to\infty$, $\mathcal{P} \to 0$, but $\epsilon$ is large because $W_{diss}$ approaches zero. In this limit, the experimentally calculated efficiency agrees with the predicted equilibrium Stirling efficiency, $\epsilon_{\infty} = \epsilon_C\left[1+\epsilon_C/\ln\left(k_{max}/k_{min}\right)\right]^{-1}$, where $\epsilon_C = 1-\frac{T_c}{T_h}$ is the Carnot efficiency (Fig. \ref{Figure2}E). In the small $\tau$ limit, both $\mathcal{P}$ and $\epsilon$ become negative because $W_{diss}$ is large. Between these two extremes, the power goes through a maximum (vertical bars in Fig. \ref{Figure2}D).

An important yardstick for any heat engine is its efficiency at maximum power, $\epsilon^*$ \cite{van2005thermodynamic,schmiedl2007efficiency,esposito2010efficiency}. For many engines under realistic operating conditions, $\epsilon^*$ agrees with the Curzon-Ahlborn (CA) efficiency, $\epsilon_{CA} = 1-\sqrt{\frac{T_c}{T_h}}$ \cite{curzon1975efficiency}. While the CA efficiency is exact for Carnot-like machines with small, and importantly, symmetric dissipation ($\Sigma_c = \Sigma_h$) in the cold and hot isotherms \cite{esposito2010efficiency}, it has proven to be a remarkably robust benchmark for both mesoscale Carnot \cite{martinez2016brownian} and Stirling engines \cite{blickle2012realization,roy2021tuning,krishnamurthy2023overcoming}. This is because, under the typical operating conditions of these engines, namely, small temperature ratios  ($T_c/T_h\lessapprox 1$) and small compression ratios ($k_{max}/k_{min} \approx O(1)$), the efficiency at maximum power is always expected to be close to $\epsilon_{CA}$, even when the dissipation is asymmetric. Our engines with $T_c/T_h\approx 0.85$, and $k_{max}/k_{min} \approx 1.5$ are no exception, with the experimental values of $\epsilon^*$ showing excellent agreement with $\epsilon_{CA}$ (Fig. \ref{Figure2}E).

Since this agreement does not reveal if the dissipation is symmetric or asymmetric, we calculated $\Sigma_c$ and $\Sigma_h$ by fitting Eqn. \ref{W_finite} to the work done along the hot
($\langle W_h\rangle$) and cold ($\langle W_c\rangle$) isotherms. The engine made of the smaller rattler exhibited symmetric dissipation ($\Sigma_c = \Sigma_h \approx 0.03k_BT^{eff}_c$, Fig. \ref{Figure3}A), while for the one with the bigger rattler, $\Sigma_c$ is almost 60$\%$ larger than $\Sigma_h$ (Supplementary Fig. 12). These findings are unexpected, as there is typically greater dissipative cost during expansion, where the particle must diffusively explore a growing volume, than during compression, where it can equilibrate rapidly since it is actively guided by the trap confining force \cite{martinez2016brownian,roy2021tuning,roy2023harnessing}.

To understand the origin of this enhanced dissipation, we tracked $\langle x^2_{cyc}(t/\tau)\rangle$ - a proxy for the instantaneous volume explored by the granule, throughout the engine cycle (Fig. \ref{Figure3}B, also see Supplementary Fig. 13). In the quasistatic limit, $\langle x^2_{cyc}(t/\tau)\rangle$ tracks the changes in $k$, and, as expected, the variance is at its minimum by the end of the compression step ($t/\tau =0.5$, $k = k_{max}$). Unexpectedly, for the fast cycles, the working medium fails to equilibrate even during compression, with $\langle x^2_{cyc}(t/\tau)\rangle$ being larger than its quasistatic value by the end of the step. This incomplete compression results in an unexpectedly large dissipation that matches or even exceeds that from expansion.

This hindrance to equilibration during compression stems from a feature of our engine design. The magnetic trap not only confines the granule laterally but also pulls it downwards, i.e., the frictional force between the particle and the bottom plate -- a `viscous-like' damping -- can increase with $k$. We determined the damping coefficient, $\gamma = k/2\pi f_c$, from Lorentzian fits to the steady-state power spectra for the measurements shown in Fig. \ref{Figure1}C and Supplementary Fig. 9. As expected, $\gamma$ is larger in the cold isotherm where the injected energy is smaller (Fig. \ref{Figure3}C and Supplementary Fig. 14). Remarkably, $\gamma$ grows steeply for large $k$ values in the cold isotherm for the smaller rattler, while for the bigger rattler, this growth is almost linear with $k$ in both isotherms, but is steeper in the cold one. This enhanced damping, arising from a non-ideal coupling between the system and the bath ($k-\gamma$ coupling), impedes equilibration during compression. 

Taken together, our work extends single-particle heat-engine phenomenology and finite-time thermodynamic constraints to a new class of engines whose working medium is an \textit{athermal}, nonlinear, driven-dissipative granule with engineered Brownian-like dynamics. Beyond reproducing the finite-time power-efficiency trade-off, our central finding is a non-ideal system-bath coupling that manifests as a control parameter-dependent drag, fundamentally altering the usual expansion–compression dissipation asymmetry expected under weak, control-independent drag operation. Such non-ideal couplings present a new knob to optimize the finite-time performance of small engines by altering where irreversibility occurs within a cycle. These couplings are not restricted to granular media; they can arise, for instance, in an engine with an optically trapped colloid near a wall, where radiation pressure alters the bead–wall separation and thus the hydrodynamic friction. 

Experimentally, our tabletop platform enables engineering the dynamics of the working medium (underdamped or overdamped), and endowing it with additional degrees of freedom, such as shape, flexibility \cite{ben2001knots}, activity \cite{arora2021emergent}, and spin \cite{liebchen2022chiral}. Engines leveraging these features, especially when connected to nonequilibrium baths, open new opportunities for energy conversion and control.

\section*{Data Availability}
All study data are included in the article or supplementary information.

\bibliography{apssamp}

\section*{Acknowledgements}
We thank Umesh Waghmare for comments on the manuscript. \textbf{N.R.} and \textbf{P.A.} thank the Jawaharlal Nehru Centre for Advanced Scientific Research, Bangalore, INDIA, for a research fellowship. \textbf{A.K.S.} thanks the Anusandhan National Research Foundation, Government of India, for the National Science Chair. \textbf{R.G.} thanks Suman Maji and Sudip Ghosh for help with the magnetic characterization studies, and the Department of Science and Technology, INDIA, for financial support through the SwarnaJayanti Fellowship Grant (DST/SJF/PSA-03/2017-22). 

\section*{Author Contributions}
\textbf{N.R.}: Methodology, Software, Validation, Formal Analysis, Investigation, Data Curation, Visualization, Writing - Review \& Editing. \textbf{P.A.}: Methodology. \textbf{A.K.S.}: Methodology, Validation, Writing - Review \& Editing. \textbf{R.G.}: Conceptualization, Methodology, Validation, Investigation, Formal Analysis, Visualization, Writing - Original Draft, Writing - Review \& Editing, Supervision, Project Administration, Funding Acquisition.

\section*{Competing Interests}
The authors declare no competing interests.
\newpage

\begin{singlespace}

\begin{figure}[tbp]
\includegraphics[width=1\textwidth]{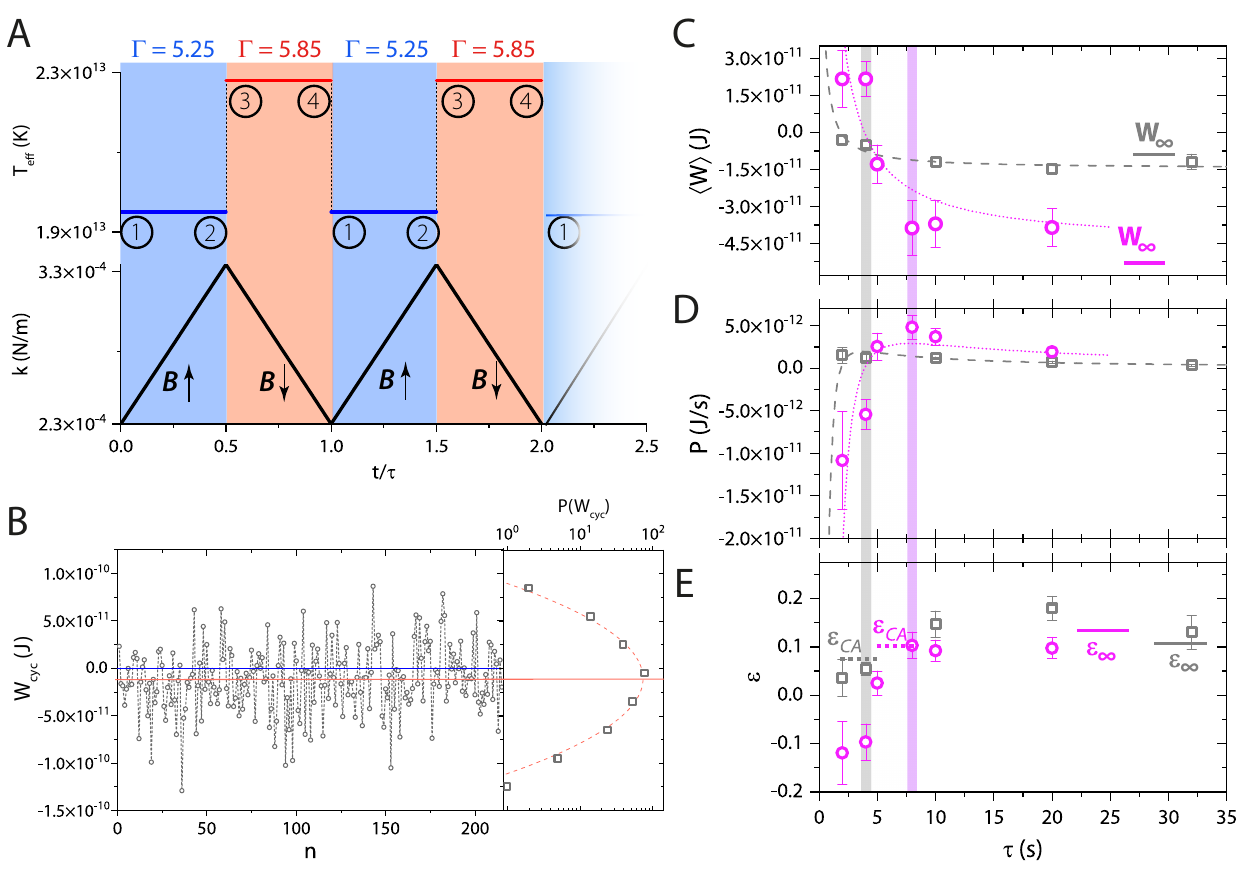}
\caption{\textbf{Energetics of a single-granule Stirling heat engine.} \textbf{(A)} Stirling cycle with the smaller rattler. \textcircled{1}$\to$\textcircled{2} Isothermal compression of the working medium by increasing trap stiffness from $k_{min} = 2.25\times10^{-4}$ N/m to $k_{max} = 3.35\times10^{-4}$ N/m at $T^{eff}_c = 1.93\times 10^{13}$ K. \textcircled{2}$\to$\textcircled{3} Isochoric heating to $T^{eff}_h = 2.3\times 10^{13}$ K at $k = k_{max}$. \textcircled{3}$\to$\textcircled{4} Isothermal expansion by decreasing the trap stiffness from $k_{max}$ to $k_{min}$ at $T^{eff}_h$. \textcircled{4}$\to$\textcircled{1} Isochoric cooling to $T^{eff}_c$ at $k = k_{min}$. The compression ratio, $k_{max}/k_{min} \approx 1.5$, and the temperature ratio $T_c^{eff}/T_h^{eff} \approx 0.85$. \textbf{(B)} Left panel: Work done per cycle, $W_{cyc}$, versus cycle number, $n$, for cycle duration $\tau = 32$ s for the smaller rattler. The red line represents the cycle-averaged work $\langle W\rangle$, which is negative. Right panel: Work distribution, $P(W_{cyc})$. The dashed line is a Gaussian fit to the data. \textbf{(C-E)} Cycle time dependence of work, power $\mathcal{P}$ and efficiency $\epsilon$ for the smaller (squares) and the larger (circles) rattler. In \textbf{(C)}, the lines represent fits to Eqn. \ref{W_finite}. In \textbf{(D)}, the lines represent fits to $\mathcal{P}(\tau) = - \frac{\langle W_{diss}\rangle + \langle W_\infty \rangle}{\tau}$. The vertical bars correspond to the cycle time at which the power peaks. The solid and dashed horizontal bars in \textbf{(C)} and \textbf{(E)} represent the predicted equilibrium Stirling work, $W_\infty$, and equilibrium efficiency, $\epsilon_\infty$, respectively, while the short dashed lines in (\textbf{E}) represent the Curzon-Ahlborn efficiency, $\epsilon_{CA}$.}
\label{Figure2}
\end{figure}

\begin{figure}[tbp]
\includegraphics[width=1\textwidth]{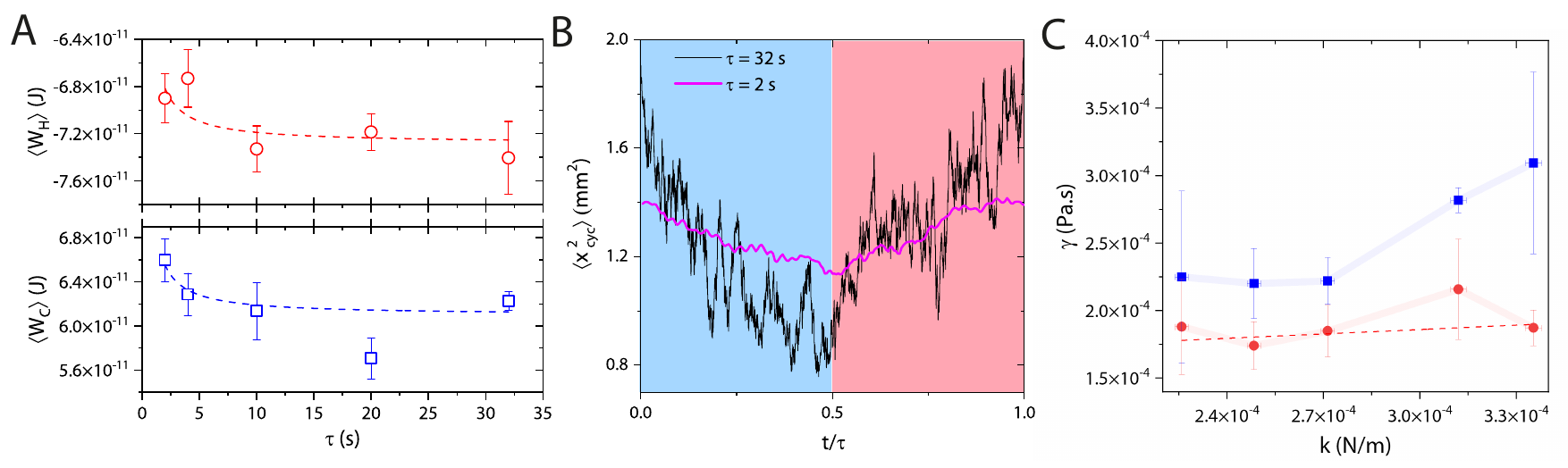}
\caption{\textbf{Stiffness-dependent damping results in enhanced dissipation in the compression stroke.} All data shown is for the smaller rattler.  \textbf{(A)} Cycle-averaged work in the hot ($\langle W_h \rangle$) and cold ($\langle W_c \rangle$) isotherms. The lines represent fits to Eqn. \ref{W_finite}. \textbf{(B)} Evolution of the cycle-averaged particle position variance, $\langle x^2_{cyc}(t/\tau)\rangle$, during cycle execution. Here, $\langle\rangle$ denotes an averaging over time instances separated by $\tau$. The pale blue and red shaded regions correspond to the compression and expansion steps, respectively. \textbf{(C)} The damping coefficient, $\gamma$, in the hot (circles) and cold (squares) isotherms, versus $k$. The dashed line is a fit to the data. The error bars in \textbf{(A)} and \textbf{(C)} denote SEM.}
\label{Figure3}
\end{figure}
\end{singlespace}
\end{document}